\documentclass{elsart}

 \usepackage{graphicx}
 \usepackage{epsfig}
\usepackage{amssymb}

\begin{document}
\title{Quantum Dating Market}
\author{O.G. Zabaleta and C.M. Arizmendi}

\address
{ Depto. de F\'{\i}sica,
Facultad de Ingenier\'{\i}a, \\
Universidad Nacional de  Mar del
Plata,\\
Av. J.B. Justo 4302, \\
7600 Mar del Plata, Argentina}

\section*{Abstract}

 We consider  the dating market decision problem under the quantum mechanics point of view.
 Quantum states whose associated amplitudes are modified by men strategies are used to represent women.
 Grover quantum search algorithm is used as a playing
strategy. Success is more frequently obtained by playing quantum than playing classic.

\section*{Introduction}

Quantum techniques, as a convenient language to generalize classical probability theory have gained attention recently. An example is provided by so-called quantum
games, introduced by Meyer \cite{Meyer1999} and Eisert et al. \cite{Eisert1999}, that allow effects which are impossible in a classical setting.

 Other examples of application of quantum techniques to classical problems are the Shor algorithm \cite{shor}, which is purely quantum-mechanical but is solving the classical factoring
problem  and the contribution of Lov Grover \cite{Boyer, Grov2}, who showed a way to speed up the search for items in an N-item database from $O(N)$ steps to
$O(\sqrt{N})$ steps. These and many other examples, show that there is no contradiction in using quantum techniques to describe
  non-quantum mechanical problems and solve hard to solve problems with classical tools. Decision theory that deals with
 decisions made under uncertain conditions by real humans is yet another problem of interest that presents both characteristics.

 Widely observed phenomena of non-commutativity
  in patterns of behavior exhibited in experiments on human decisions and choices cannot be obtained with classical
  decision theory \cite{Lambert-Mogiliansky} but can be adequately described by putting quantum mechanics and decision theory together.
Quantum mechanics and decision theory have been  recently combined \cite{Lambert-Mogiliansky,Sornette, Temzelides}
 to take into account the indeterminacy of preferences that
  are determined  only when the action takes place. An agent is described by a state that is a superposition of potential preferences
  to be projected onto one of the possible behaviors at the time of the interaction. In addition to the main goal of modeling uncertainty of
  preferences that is not due to lack of information, this formalism seems to be adequate to describe widely observed phenomena of non-commutativity
  in patterns of behavior.

 Within this framework, we study
the dating market decision problem that takes into account progressive mutual learning. This problem is a variation on the Stable Marriage Problem introduced by
Gale and Shapley almost four decades ago \cite{GaleShapley}, that has been recently reformulated in a partial information approach \cite{Zhang,Zhang1}. The dating
market problem may be included in a more general category of matching problems where the elements of two sets have to be matched by pairs. Matching problems have
broad implications in economic and social contexts \cite{Roth,Mendez}. As possible applications one could think of job seekers and employers, lodgers and landlords,
men and women who want to date \cite{Das,ariz}, or solitary ciliates {\sl courtship rituals} \cite{ciliates}. In our model players have a list of preferred partners
on the other set. Quantum exploration of partners is compared with classical exploration at the dating set. Nevertheless dating is not just finding, but also being
accepted by the partner. The preferences of the chosen partner are important in quantum and classic performances.

\section*{The quantum dating game}

In the classic dating market game \cite{Das, ariz}, men choose women simultaneously from $N$ options, looking for those women who would have some ``property" they
want. Unlike the traditional game, in the quantum version of the dating game, players get the chance to use quantum techniques, for example they can explore their
possibilities using a quantum search algorithm. Grover algorithm capitalizes quantum states superposition characteristic to find some ``marked" state from a group
of possible solutions in considerably less time than a classical algorithm can do \cite{Grov1}. That state space must be capable of being translatable, say to a
graph $G$ where to find some particular state which has a searched feature or distinctive mark, throughout the execution of the algorithm. By ``distinctive mark" we
mean problems whose algorithmic solution are inspired by physical processes. Furthermore it is possible to guarantee that the searched node is marked by a minimum
(maximum) value of a physical property included in the algorithm.

Let agents be coded as Hilbert space base states. As a result, men are able to choose from $N_w$ women set $W=\{|0\rangle, |1\rangle, ..., |N_w-1\rangle \}$. Table
1 displays four women states in the first column and some feature that makes them unique in the second column which we will code with a letter for simplicity.

\begin{table}[ht]
  \centering
  \begin{tabular}{|c|c|}
    \hline
    woman & feature \\
    \hline
    $|0\rangle$ & a \\
    $|1\rangle$ & b \\
    $|2\rangle$ & c \\
    $|3\rangle$ & d \\
    \hline
  \end{tabular}
  \caption{Sample woman database. Left column contains women states and right column displays a letter representing some feature or a feature set
  that characterizes each woman on the left.}\label{wt}
\end{table}
If a player is looking for a woman with a feature ``d", the table must be searched on its second column and when the desired ``d" is found, look at the first column
where the corresponding chosen woman state is: $|3\rangle$ in this example. The procedure is very simple if the table has just a few rows, but when the database
gets bigger, the table in the best case would have to be entered  $N_w/2$ times \cite{Qlearn, Roman}. Under this framework we propose to use Grover algorithm in
order to achieve man's decision in less time. Without losing generality let $N_w=2^n$ being $n$ the qubits needed to code $N_w$ women. Quantum states transformation
are made by applying Hilbert space operators $U$ to them, following $\Psi_1=U_1\Psi_0$ is a new system state starting from $\Psi_0$. As a consequence any quantum
algorithm can be thought as a set of suitable linear transformations.  Grover algorithm starts with $n$ qubits in $|0\rangle$, resulting
$\psi_{ini}=|00..00\rangle\equiv|0\rangle^{\bigotimes n}$ the system initial state, where $\bigotimes$ symbol denotes Kronecker tensor product. Initially, the woman
identified by state $|0\rangle$ is chosen with probability one. The next step is to create superposition states and like many other quantum algorithms Grover uses
Hadamard transform to do this task since it maps $n$ qubits initialized with $|0\rangle$ to a superposition of all $n$ orthogonal states in the $|0\rangle$,
$|1\rangle$,.. $|n-1\rangle$ basis with equal weight, $\psi_{1}= H\psi_{ini}=\frac{1}{\sqrt{N_w}}\sum^{N_w-1}_{i=0}|i\rangle$. One-qubit  Hadamard transform matrix
representation is (\ref{Had}), and n-qubits extension is $H^{\bigotimes n}$, see \cite{NielChuang},

\begin{equation}\label{Had}
    H=\frac{1}{\sqrt{2}}\left(%
\begin{array}{cc}
  1 & 1 \\
  1 & -1 \\
\end{array}%
\right)
\end{equation}

Another quantum search algorithms characteristic, is the ``Oracle", which is basically a black box capable of marking the problem solution. We call $U_f$ the
operator which implement the oracle
\begin{equation}\label{Uf}
    U_f(|w\rangle |q\rangle)=|w\rangle|q\oplus f(w)\rangle,
\end{equation}
where $f(w)$ is the oracle function which takes the value $1$ if $w$ correspond to the searched woman, $f(w)=1$, and if it is not the case it takes the value $0$,
$f(w)=0$. The value of $f(w)$ on a superposition of every possible input $w$  may be obtained \cite{NielChuang}. The algorithm sets the target qubit $|q\rangle$ to
$\frac{1}{\sqrt{2}}(|0\rangle-|1\rangle)$. As a result, the corresponding mathematical expression is:
\begin{equation}\label{or}
   |w\rangle(\frac{|0\rangle-|1\rangle}{\sqrt{2}})\longmapsto^{U_f}(-1)^{f(w)}|w\rangle(\frac{|0\rangle-|1\rangle}{\sqrt{2}})
\end{equation}
Observe that the second register is in an eigenstate, so we can ignore it, considering only the effect on the first register.

\begin{equation}\label{or2}
   |w\rangle\longmapsto^{U_f}(-1)^{f(w)}|w\rangle
\end{equation}

Consequently, if $f(w)=1$  a phase shift is produced, otherwise nothing happens. As we already stated our algorithm is based on the classical Gale-Shapley (GS)
algorithm \cite{GaleShapley} which assigns the role of proposers to the elements of one set, the men say, and of judges to the elements of the other.

Actually, for a more symmetric formulation of the algorithm where both sets are, at the same time, proposers and judges,
 it would be necessary another oracle
which evaluates women features matching by means of another function $g(x)$ \cite{Pang}, but we will not go into that. As far as we are concerned up to now the
Oracle is a device capable of recognizing and ``mark" a woman who has some special feature, said hair color, money, good manners, etc. Oracle operator $U_f$ makes
one of two central operations comprising of a whole operation named Grover iterate $G$ (Fig.1), and a rotation operator $U_R$, or conditional phase shift operator
represented by equation (\ref{ph}).
\begin{figure}[h]\label{g1} {\psfig{file=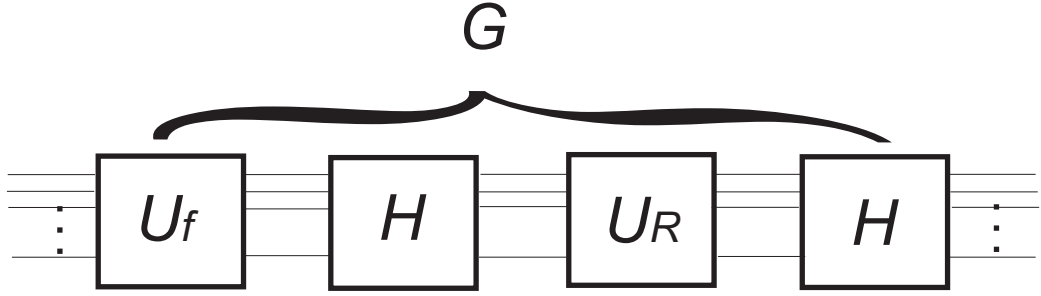}}
\caption{The Grover Iterate}
\end{figure}

 $U_R$ and
$U_f$, together with Hadamard transformations represented by $H$ blocks (\ref{Had}), in the order depicted by (Fig. 1), make the initial state vector asymptotically
going to reach the solution state vector amplitudes. The symbol $I$ in $U_R$ equation is the identity operator.
\begin{equation}\label{ph} U_{R}=2|0\rangle\langle 0|-I
\end{equation}Furthermore, after applying
Grover iterate, $G$, $O(\sqrt{N_w})$ times, the man finds the woman he is looking for. In Figure 1 Grover iterate is shown and Grover quantum algorithm scheme is
depicted in Figure 2.

\begin{figure}[h]
{\psfig{file=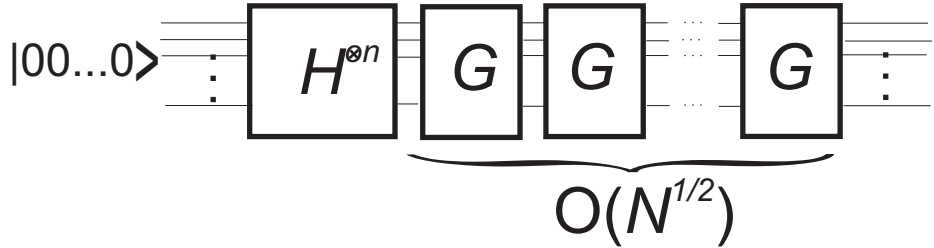}} \caption{Grover Quantum searching algorithm}
\end{figure}\label{gsch}

As the number of iterations the algorithm makes depends on the size of the options set, this must be known at the beginning of simulations. Every operator has its
matrix representation to be used in simulations. We suppose the player chooses a woman who has some specific particularity that would distinguish her from any other
of the group, so we construct matrix $U_{f}$ and other matrixes for that purpose. The evolution of the squared amplitude with the iteration number is shown in
Figure 3. The searched state amplitude is initially the same for all possible states $|i>$ in the $\Psi_1$ expression. The fast increasing of the probability to
find the preferred state on each iteration contrasts with the decreasing of the probability to find every other state. The example displayed is for $N_w=1024$ women
and as the can be seen in Figure 3, the number of iterations needed to get certainty to find the preferred woman are $25$. Classically, a statistical algorithm
would need approximately $N_w=1024$ iterations.
\begin{figure}[h]\label{Gasymp}
\centering
 {\psfig{file=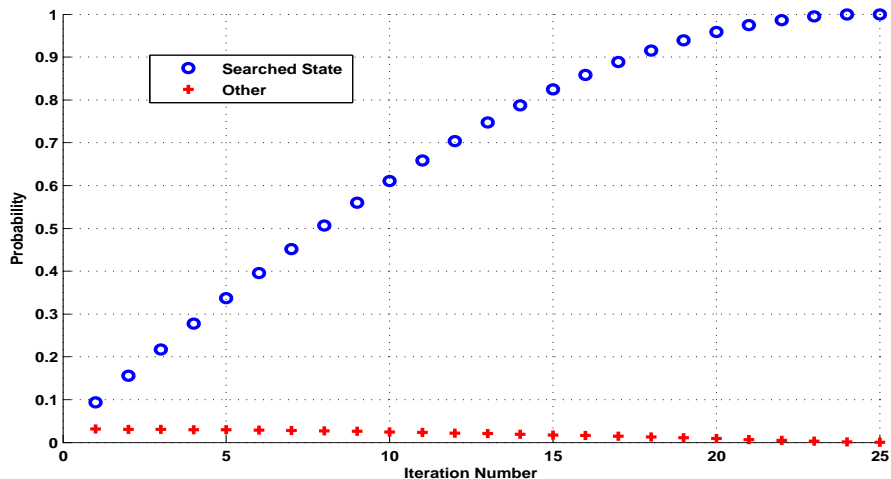,width=14 cm, height=7 cm}}
\caption{Evolution of the probability to find the chosen woman and the probability to find other woman as a function of the iteration number with Grover's
algorithm.}
\end{figure}

Thus when a given man who wants to date a $N_w$ size set selected woman, he must set his own $U_f$ operator out, according to his preferences, and then let the
algorithm do the job. The case of $N_m$ men may be obtained generalizing the single man case: every one of them must follow the same steps. Nevertheless, achieving
top choice is hard because of competition from other players and your dream partner may not share your feelings. If all players play quantum, the time to find woman
is not an issue and the $N$ stable solutions will be the same  as for the classic formulation \cite{Omero}.

\section*{Quantum vs classic}

To compare  the quantum approach efficiency with the classical one we will consider  some players playing quantum and others playing classic. Let us follow the
evolution of two agents representative from each group, $Q$ and $C$ respectively.

$Q$, that plays quantum can keep his state as a linear combination of all the prospective results when unitary transforms such as the described above for Grover's
algorithm are applied, provided no measurement producing collapse to any of them is done.
 On the other hand,  the only way $C$ has to
search such a database is to test the elements sequentially against the condition until the target is found. For a database of size $N$, this brute force search
requires an average of $O(N/2)$ comparisons \cite{Boyer}.

In order to compare performances two different games where both men want to date with the same woman are presented: In the first one player $Q$ gives player $C$ the
chance to play first and both have only one attempt per turn, which means only one question to the oracle. The second game, in order that $Q$ plays handicapped, is
set out in the way that $C$ can play $N/2$ times while $Q$ only once, and player $C$ plays first again. The classic player $C$ plays without memory of his previous
result and therefore, in every try he has $1/N$ probability to find the chosen woman to date. It is important to remark that only one Grover iterate is performed by
$Q$ (Figure 1) in both games.

The player who invites the chosen woman first has more chances to succeed, as well as that who asks the same woman more times. Nevertheless the woman has the last
word, and therefore the dating success for each player depends on that woman preferences. So, let us define $P^i_c$ as the probability that woman $i$ accepts dating
the classic player $C$ and $P^i_q$ as the probability that she accepts the quantum player $Q$ proposal. In order to compare performances, we consider  $T=1000$
playing times on turns and count the dating success times, then calculate the mean relative difference between $Q$ and $C$ success total number as $D/T=\frac{Q
success - C success}{T}$, for different woman acceptation probabilities.

Initially, both players begin with the  system in the initial state $\psi_{1}=\frac{1}{\sqrt{N_w}}\sum^{N_w-1}_{i=0}|i\rangle$, therefore the probability to select
any woman is the same for both, $p(w_i)=1/N$. In the next step the Oracle marks one of the prospective women state according men preferences.

The results are highly dependent on the women set size  $N$  because, as mentioned above, Grover algorithm needs $O(\sqrt(N))$ steps to find the quantum player's
chosen partner while the classic player must use $O(N)$ for the same task. In the case of only one woman and one man, for example, classic and quantum will not have
any advantage on searching and the dating success difference for the first game will depend only on that woman preferences, that is, if $P_c > P_q$ then $D/T < 0$
and  the quantum player will do better when $P_q > P_c$ . Similar chances for both players is not usual in most quantum games, such as, for example the coin flip
game introduced by Meyer \cite{Meyer1999} where the quantum player always beats the classic player in a {\sl ``mano a mano"} game. For a two women set $Q$ uses only
one step, but $C$ needs two steps to find the right partner. In this case $Q$ does better when $P_q > P_c/4$. Winning conditions improve significantly for the
quantum player for increasing $N$, but not in a monotonous way, because the probability to find the chosen partner in one step is associated with the projection of
the state obtained with only one  Grover iterate (Fig. 1)  on the corresponding chosen partner state ( In Fig. 3 this probability is $ \approx 0.1$) for the quantum
player while the classic player odds to meet his partner is always $1/N$. On the other hand for the second game, the classic player has $N/2$ opportunities while
$Q$ has, as for the first game, only one step. Therefore the advantage increases for $C$ in this game for increasing $N$.

In order to facilitate comprehension the set size in the simulations results shown is $N=8$.

Under the first game conditions both players have only one attempt by turn. Since $C$ cannot modify state $\psi_{1}$ amplitudes, he has $1/8$ chance to be right. On
the other hand player $Q$, using Grover algorithm as his strategy, can modify states amplitudes in order to increase his chances to win, reaching $0.78$ as the
probability to find his preferred woman in only one iteration. Figure 4 shows that situation outcomes for different $P^i_c$ and $P^i_q$ combinations. The vertical
axis depicts $D/T$ values as a function of $P^i_c$ and $P^i_q$ respectively. $D/T$ is positive for all $P^i_c$ and $P^i_q$ values used in the simulation, which
means that even at extremes where $P^i_c >> P^i_q$, the quantum player performs better. However there is a very  small region where $P^i_c \approx 1$ and $P^i_q
\approx 0$ not shown in the figure that corresponds to a prevailing $C$.

\begin{figure}[h]\label{ClaQN1}
\centering
 {\psfig{file=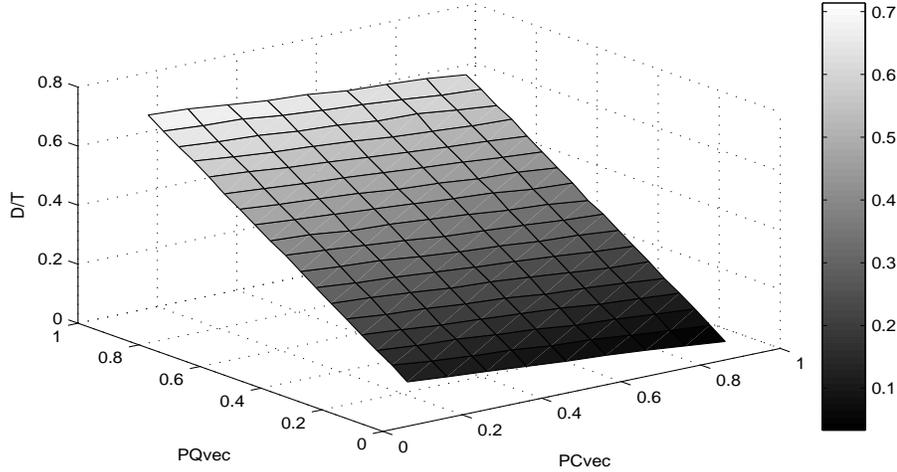,width=14 cm, height=7 cm}}
\caption{First game: One attempt for both players. Mean relative difference between $Q$ and $C$ success total number as $D/T=\frac{Q success - C success}{T}$, for
different woman acceptation probabilities $P^i_c$ and $P^i_q$. $Q$ outperforms $C$ in all shown cases. The small region where $C$ prevails is not shown.}
\end{figure}

Under the second game conditions player $C$ have $\frac{N}{2}=4$ attempts before $Q$ plays. After each  $C$ attempt the system is forced to collapse to one base
state, so a third party, that could be the oracle, arrange the states again and mark the solution. As we explained above, to mark a state means to change its phase
but nothing happens to the state amplitude, consequently, for the classic player $C$, the probability that state results the one the Oracle have signaled is, marked
or not,
 $1/N = 1/8$, even though, due to his ``insistence", he tries $\frac{N}{2}=4$
times, his dating success chances increase considerably with respect to the first case. Figure 5 shows the corresponding results, where it is possible to see that
classic player $C$ begins to outperform $Q$ when $P^i_c >> P^i_q$, that is, when woman has a marked preference for player $C$.

Player $C$ probability to find the chosen woman can increase to $\frac{1}{2}$ when using a classical algorithm like ``Brute-Force algorithm".  As shown in figure 5,
when $C$ has $\frac{N}{2}=4$ tries while $Q$ has only one, $C$'s odds of success in dating increases, and there are zones on the graph where $D/T < 0$. This implies
that player $C$ outperforms player $Q$. Nevertheless, to achieve that, the chosen woman preferences must be considerably greater for the classic player, that is
$P^i_c > 2 P^i_q$.

\begin{figure}[h]\label{ClaQN4}
\centering
 {\psfig{file=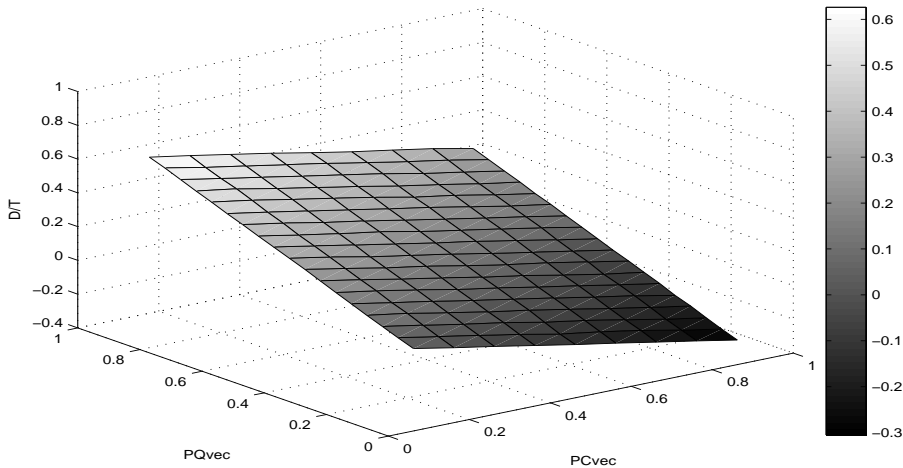,width=14 cm, height=7 cm}}
\caption{Second game: Classic player $C$ has four tries while $Q$ has only one. Mean relative difference between $Q$ and $C$ success total number as $D/T=\frac{Q
success - C success}{T}$, for different woman acceptation probabilities $P^i_c$ and $P^i_q$.  $C$ outperforms $Q$ when $P^i_c >> P^i_q$}
\end{figure}

\section*{Conclusion}
We have introduced a quantum formulation for decision matching problems, specifically
 for the dating game. In that framework women are
represented with quantum states whose associated amplitudes must be modified by men's selection strategies, in order to increase a particular state amplitude and to
decrease the others, with the final purpose to achieve the best possible choice when the game finishes. This is a highly time consuming task that takes a $O(N)$
runtime for a classical probabilistic algorithm, being $N$ the women database size. Grover quantum search algorithm is used as a playing strategy that takes the man
$O(\sqrt{N})$ runtime to find his chosen partner. As a consequence, if every man uses quantum strategy, no one does better than the others, and stability is quickly
obtained.

The performances of quantum vs. classic players depend on the number of players $N$. In a ``one on one'' game there is no advantage from any of them and the woman
preferences rule. Similar chances for quantum and classic players in ``one on one'' situation is not usual in most quantum games, such as Meyer's penny flip
\cite{Meyer1999}. Winning conditions improve for the quantum player for increasing $N$ and the same number of attempts, but not in a monotonous way.  If the game is
set in order that the classic player
 has $\frac{N}{2}$ opportunities and the quantum player only one, the former player begins to have an advantage over
 the quantum one when his probability to be accepted by the chosen woman is much higher than the probability for the
 quantum player for small $N$. The advantage increases for $C$ in this game for increasing $N$ because the opportunities for $C$ are $N/2$ while $Q$ has only one.
 As quantum entanglement enhances the ``speed'' of evolution of certain quantum states \cite{Plastino}
 in future analysis we will introduce entanglement between players in order to see if it provides any advantage to
 entangled players and changes the stable solutions.

\section{Acknowledgments}

The authors thank K. I. Mazzitello for helpful discussions on dating problems. This work was partially supported by Universidad Nacional de Mar del Plata and ANPCyT
(PICT07-00807).

\end{document}